\documentclass[structabstract]{aa_customized}

\usepackage{amsmath}   
\usepackage{graphicx}
\usepackage{txfonts}
\usepackage{natbib}
\usepackage[e]{esvect}  
\usepackage{wasysym}   

\bibpunct{(}{)}{;}{a}{}{,} 

\usepackage[pdftitle={w}, colorlinks = true, breaklinks = true, citecolor = blue, linkcolor = blue, urlcolor = blue, pdfauthor = {Ren\'{e} Heller}]{hyperref}

\def\wu#1{\sqrt{{#1}\!\;\,}^{\!\;\!\rule[+1.371ex]{.05em}{.6ex}}\;\!}

\begin{document}

\title{Runaway greenhouse effect on exomoons due to\\ irradiation from hot, young giant planets}


\author{R. Heller\inst{1}  \and R. Barnes \inst{2,3} }

\institute{
McMaster University, Department of Physics and Astronomy, 1280 Main Street West, Hamilton (ON) L8S 4M1, Canada\\ \href{mailto:rheller@physics.mcmaster.ca}{rheller@physics.mcmaster.ca}
\and
University of Washington, Department of Astronomy, Seattle, WA 98195, USA\\
\href{mailto:rory@astro.washington.edu}{rory@astro.washington.edu}
\and
Virtual Planetary Laboratory, USA
}

\titlerunning{Runaway greenhouse effect on exomoons due to planetary irradiation}

\authorrunning{Heller \& Barnes}

\date{submitted June 15, 2013 \ -- \ revision \#1 September 6, 2013 \ -- \ revision \#2 October 17, 2013 \ -- \ accepted November 1, 2013}

\abstract
{The \textit{Kepler} space telescope has proven capable of detecting transits of objects almost as small as the Earth's Moon. Some studies suggest that moons as small as $0.2$ Earth masses can be detected in the \textit{Kepler} data by transit timing variations and transit duration variations of their host planets. If such massive moons exist around giant planets in the stellar habitable zone (HZ), then they could serve as habitats for extraterrestrial life.}
{While earlier studies on exomoon habitability assumed the host planet to be in thermal equilibrium with the absorbed stellar flux, we here extend this concept by including the planetary luminosity from evolutionary shrinking. Our aim is to assess the danger of exomoons to be in a runaway greenhouse state due to extensive heating from the planet.}
{We apply pre-computed evolution tracks for giant planets to calculate the incident planetary radiation on the moon as a function of time. Added to the stellar flux, the total illumination yields constraints on a moon's habitability. Ultimately, we include tidal heating to evaluate a moon's energy budget. We use a semi-analytical formula to parametrize the critical flux for the moon to experience a runaway greenhouse effect.}
{Planetary illumination from a 13-Jupiter-mass planet onto an Earth-sized moon at a distance of ten Jupiter radii can drive a runaway greenhouse state on the moon for about 200\,Myr. When stellar illumination equivalent to that received by Earth from the Sun is added, then the runaway greenhouse holds for about 500\,Myr. After 1000\,Myr, the planet's habitable edge has moved inward to about 6 Jupiter radii. Exomoons in orbits with eccentricities of 0.1 experience strong tidal heating; they must orbit a 13-Jupiter-mass host beyond 29 or 18 Jupiter radii after 100\,Myr (at the inner and outer boundaries of the stellar HZ, respectively), and beyond 13 Jupiter radii (in both cases) after 1000\,Myr to be habitable.}
{If a roughly Earth-sized, Earth-mass moon would be detected in orbit around a giant planet, and if the planet-moon duet would orbit in the stellar HZ, then it will be crucial to recover the orbital history of the moon. If, for example, such a moon around a $13$-Jupiter-mass planet has been closer than $20$ Jupiter radii to its host during the first few hundred million years at least, then it might have lost substantial amounts of its initial water reservoir and be uninhabitable today.}

\keywords{Astrobiology -- Celestial mechanics -- Planets and satellites: general -- Radiation mechanisms: general}

\maketitle

\section{Introduction}
\label{sec:introduction}

The advent of exoplanet science in the last two decades has led to the compelling idea that it could be possible to detect a moon orbiting a planet outside the solar system. By observational selection effects, such a finding would reveal a massive moon, because its signature in the data would be most apparent. While the most massive moon in the solar system -- Jupiter's satellite Ganymede -- has a mass roughly $1/40$ the mass of Earth ($M_\oplus$), a detectable exomoon would have at least twice the mass of Mars, that is, $1/5\,M_\oplus$ \citep{2009MNRAS.400..398K}. Should these relatively massive moons exist, then they could be habitats for extrasolar life.

One possible detection method relies on measurements of the transit timing variations (TTV) of the host planet as it periodically crosses the stellar disk \citep{1999A&AS..134..553S,2007A&A...470..727S,2009MNRAS.392..181K,2013MNRAS.430.1473L}. To ultimately pin down a satellite's mass and its orbital semi-major axis around its host planet ($a_\mathrm{ps}$), it would also be necessary to measure the transit duration variation \citep[TDV,][]{2009MNRAS.392..181K,2009MNRAS.396.1797K}. As shown by \citet{2013MNRAS.432.2549A}, \textit{Kepler}'s ability to find an exomoon is crucially determined by its ability to discern the TDV signal, as it is typically weaker than the TTV signature. Using TTV and TDV observations together, it should be possible to detect moons as small as $0.2\,M_\oplus$ \citep{2009MNRAS.400..398K}.

Alternatively, it could even be possible to observe the direct transits of large moons \citep{2006A&A...450..395S,2011ApJ...743...97T,2011MNRAS.416..689K}, as the discovery of the sub-Mercury-sized planet Kepler-37b by \citet{2013Natur.494..452B} recently demonstrated. Now that targeted searches for extrasolar moons are underway \citep{2012ApJ...750..115K,0004-637X-770-2-101,2013ApJ...777..134K} and the detection of a roughly Earth-mass moon in the stellar habitable zone \citep[][HZ in the following]{1964QB54.D63.......,1993Icar..101..108K} has become possible, we naturally wonder about the conditions that determine their habitability. Indeed, the search for spectroscopic biosignatures in the atmospheres of exomoons will hardly be possible in the near future, because the moon's transmission spectrum would need to be separated from that of the planet  \citep{2010ApJ...712L.125K}. But the possible detection of radio emission from intelligent species on exomoons still allows the hypothesis of life on exomoons to be tested.

The idea of habitable moons has been put forward by \citet{1987AdSpR...7..125R} and \citet{1997Natur.385..234W}. Both studies concluded that tidal heating can be a key energy source if a moon orbits its planet in a close, eccentric orbit \citep[for tidal heating in exomoons see also][]{2006ApJ...648.1196S,2007ApJ...668L.167D,2009ApJ...704.1341C,2009ApJ...707.1000H,2012A&A...545L...8H,2013AsBio..13...18H}. Reflected stellar light from the planet and the planet's own thermal emission can play an additional role in a moon's energy flux budget \citep{2013AsBio..13...18H,2013ApJ...774...27H}. Having said that, eclipses occur frequently in close satellite orbits which are coplanar to the circumstellar orbit. These occultations can significantly reduce the average stellar flux on a moon \citep{2012A&A...545L...8H}, thereby affecting its climate \citep{2013MNRAS.432.2994F}. Beyond that, the magnetic environment of exomoons will affect their habitability \citep{2013ApJ...776L..33H}.

Here, we investigate another effect on a moon's global energy flux. So far, no study focused on the impact of radial shrinking of a gaseous giant planet and the accompanying thermal illumination of its potentially habitable moons.\footnote{We include a term $W_\mathrm{p}$ for an additional source of illumination from the planet onto the moon in our orbit-averaged Eq.~(B1) in \citet{2013AsBio..13...18H}. This term can be attributed to heating form the planet. In \citet{2013ApJ...776L..33H}, we use methods developed here, but illumination from Neptune-, Saturn-, and Jupiter-like planets is weak.} As a giant planet converts gravitational energy into heat \citep{2003A&A...402..701B,2013NatGe...6..347L}, it may irradiate a putative Earth-like moon to an extent that makes the satellite subject to a runaway greenhouse effect. Atmospheric mass loss models suggest that desiccation of an Earth-sized planet (or, in our case, of a moon) in a runaway greenhouse state occurs as fast as within $100$ million years (Myr). Depending on the planet's surface gravitation, initial water content, and stellar XUV irradiation in the high atmosphere \citep[][see Sect. 2 and Appendix B therein]{2013AsBio..13..225B}, this duration can vary substantially. But as water loss is a complex process -- not to forget the possible storage of substantial amounts of water in the silicate mantle, the history of volcanic outgassing, and possible redelivery of water by late impacts \citep{2013oepa.book.....L} -- we here consider moons in a runaway greenhouse state to be temporarily uninhabitable, rather than desiccated forever.

\section{Methods}
\label{sec:methods}

In the following, we consider a range of hypothetical planet-moon binaries orbiting a Sun-like star during different epochs of the system's life time. As models for in-situ formation of exomoons predict that more massive host planets will develop more massive satellites \citep{2006Natur.441..834C,2010ApJ...714.1052S,2012ApJ...753...60O}, we focus on the most massive host planets that can possibly exist, that is, Jovian planets of roughly $13$ Jupiter masses.\footnote{Given only the mass, such an object might well be a brown dwarf and not a planet. We here \textit{assume} that this hypothetical object is a giant planet and that it can be described by the \citet{2003A&A...402..701B} evolution models (see below).} Following \citet{2006Natur.441..834C} and \citet{HellerReview}, such a massive planet can grow Mars- to Earth-sized moons in its circumplanetary disk.

\begin{figure}[t]
  \centering
  \scalebox{0.57}{\includegraphics{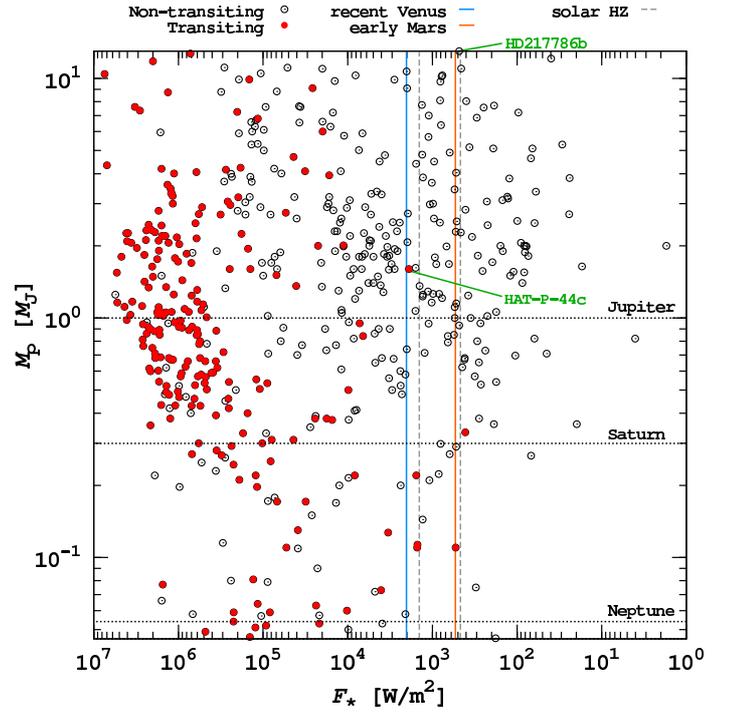}}
  \caption{Planetary masses as a function of orbit-averaged stellar illumination for objects with masses between that of Uranus and 13 times that of Jupiter. Non-transiting objects are shown as open circles; red circles correspond to transiting objects, which could allow for the detection of exomoons via TTV or TDV techniques. HD\,217786\,b is labeled because it could host Mars- to Earth-sized moons in the stellar HZ. HAT-P-44c is the most massive transiting planet in the recent Venus/early Mars HZ, whose width independent of the host star is denoted by the solid vertical lines. The dashed vertical lines denote the inner and outer HZ borders of the solar system \citep{2013ApJ...765..131K}. Obviously, some ten super-Jovian planets reside in or close to the stellar HZs of their stars, but most of which are not known to transit their stars.}
  \label{fig:planets}
\end{figure}

In Fig.~\ref{fig:planets}, we show the detections of stellar companions with masses between that of Uranus and $13\,M_\mathrm{J}$, where $M_\mathrm{J}$ denotes Jupiter's mass.\footnote{Data from \url{www.exoplanet.eu} as of Sep. 25, 2013.} While the ordinate measures planetary mass ($M_\mathrm{p}$), the abscissa depicts orbit-averaged stellar illumination received by the planet, which we compute via 

\begin{equation}
F_\star = \frac{\sigma_\mathrm{SB} T_\star^4}{\wu{1-e_\mathrm{{\star}p}^2}} \left(\frac{R_\star}{a_\mathrm{{\star}p}}\right)^2 \ .
\end{equation}

\noindent
Here, $\sigma_\mathrm{SB}$ is the Stefan-Boltzmann constant, $T_\star$ the stellar effective temperature, $R_\star$ the stellar radius, $e_\mathrm{{\star}p}$ the orbital eccentricity of the star-planet system, and $a_\mathrm{{\star}p}$ the planet's orbital semi-major axis.

Dashed vertical lines illustrate the inner and outer boundaries of the solar HZ, which \citet{2013ApJ...765..131K} locate at $1.0512\times\,S_\mathrm{eff,\odot}$ for the runaway greenhouse effect and at $0.3438\times\,S_\mathrm{eff,\odot}$ for the maximum possible greenhouse effect ($S_\mathrm{eff,\odot}=1360\,\mathrm{W\,m}^{-2}$ being the solar constant). For stars other than the Sun or planetary atmospheres other than that of Earth, the HZ limits can be located at different flux levels. Planets shown in this plot orbit a range of stars, most of which are on the main-sequence. The two solid vertical lines denote a flux interval which is between the recent Venus and early Mars HZ boundaries independent of stellar type, namely between $1.487\,{\times}\,S_\mathrm{eff,\odot}$ (recent Venus) and $0.393\,{\times}\,S_\mathrm{eff,\odot}$ (early Mars) \citep{2013ApJ...765..131K}. The purpose of Fig.~\ref{fig:planets} is to confirm that super-Jovian planets (and possibly their massive moons) exist in or near the HZ of main-sequence stars in general, rather than to explicitly identify giant planets in the HZ of their respective host stars. Yet, the position of HD\,217786\,b is indicated in Fig.~\ref{fig:planets}, because it roughly corresponds to our hypothetical test planet in terms of orbit-averaged stellar illumination at the outer HZ boundary ($483\,\mathrm{W\,m}^{-2}=0.355\,\times\,S_\mathrm{eff,\odot}$) and planetary mass \citep[$M_\mathrm{p}=13\,\pm0.8\,M_\mathrm{J}$,][]{2011A&A...527A..63M}. We also highlight HAT-P-44c because it is the most massive transiting planet in the recent Venus/early Mars HZ \citep{2013arXiv1308.2937H} and could thus allow for TTV and TDV detections of its massive moons if they exist.

\subsection{Exomoon illumination from a star and a cooling planet}

The electromagnetic spectrum of a main-sequence star is very different from that of a giant planet. M, K, and G dwarf stars have effective temperatures between $\approx3000$\,K and $\approx6000$\,K, whereas planets typically are cooler than $1000$\,K. Earth-like planets near the inner edge of the HZ around Sun-like stars are thought to have bolometric albedos as low as 0.18, increasing up to about 0.45 towards the outer HZ edge \citep{2013ApJ...765..131K}. The inner edge value would likely be significantly higher if clouds were included in the calculation \citep{2013ApJ...771L..45Y}. As a proxy, we here assume an optical albedo $A_\mathrm{s,opt}=0.3$, while an Earth-like planet orbiting a cool star with an effective temperature of $2500$\,K has an infrared albedo $A_\mathrm{s,IR}=0.05$ \citep{1993Icar..101..108K,2007A&A...476.1373S,2013ApJ...765..131K}\footnote{Strong infrared absorption of gaseous H$_2$O and CO$_2$ in the atmospheres of terrestrial worlds and the absence of Rayleigh scattering could further decrease $A_\mathrm{s,IR}$. These effects, however, would tend to warm the stratosphere rather than the surface. Stratospheric warming could then be re-radiated to space without contributing much to warming the surface.}. We therefore use a dualpass band to calculate the total illumination absorbed by the satellite

\begin{equation}\label{eq:F_i}
F_\mathrm{i} = \frac{L_*(t)(1-A_\mathrm{s,opt})}{16{\pi}a_\mathrm{*p}^2} + \frac{L_\mathrm{p}(t)(1-A_\mathrm{s,IR})}{16{\pi}a_\mathrm{ps}^2} \ \ ,
\end{equation}

\noindent
where the first term on the right side of the equation describes the stellar flux absorbed by the moon, and the second term denotes the absorbed illumination from the planet. $L_*(t)$ and $L_\mathrm{p}(t)$ are the stellar and planetary bolometric luminosities, respectively, while $a_\mathrm{*p}$ is the orbital semi-major axes of the planet around the star.\footnote{Equation~\eqref{eq:F_i} implies that the planet is much more massive than the moon, so that the planet-moon barycenter coincides with the planetary center of mass. It is also assumed that $a_\mathrm{ps}{\ll}a_\mathrm{*p}$ and that both orbits are circular.} The factor 16 in the denominators indicates that we assume effective redistribution of both stellar and planetary irradiation over the moon's surface \citep[see Sect. 2.1 in][]{2007A&A...476.1373S}. Stellar reflected light from the planet is neglected \citep[for a description see][]{2013AsBio..13...18H}.

\begin{figure}[t]
\vspace{-0.1cm}
  \centering
  \scalebox{0.591}{\includegraphics{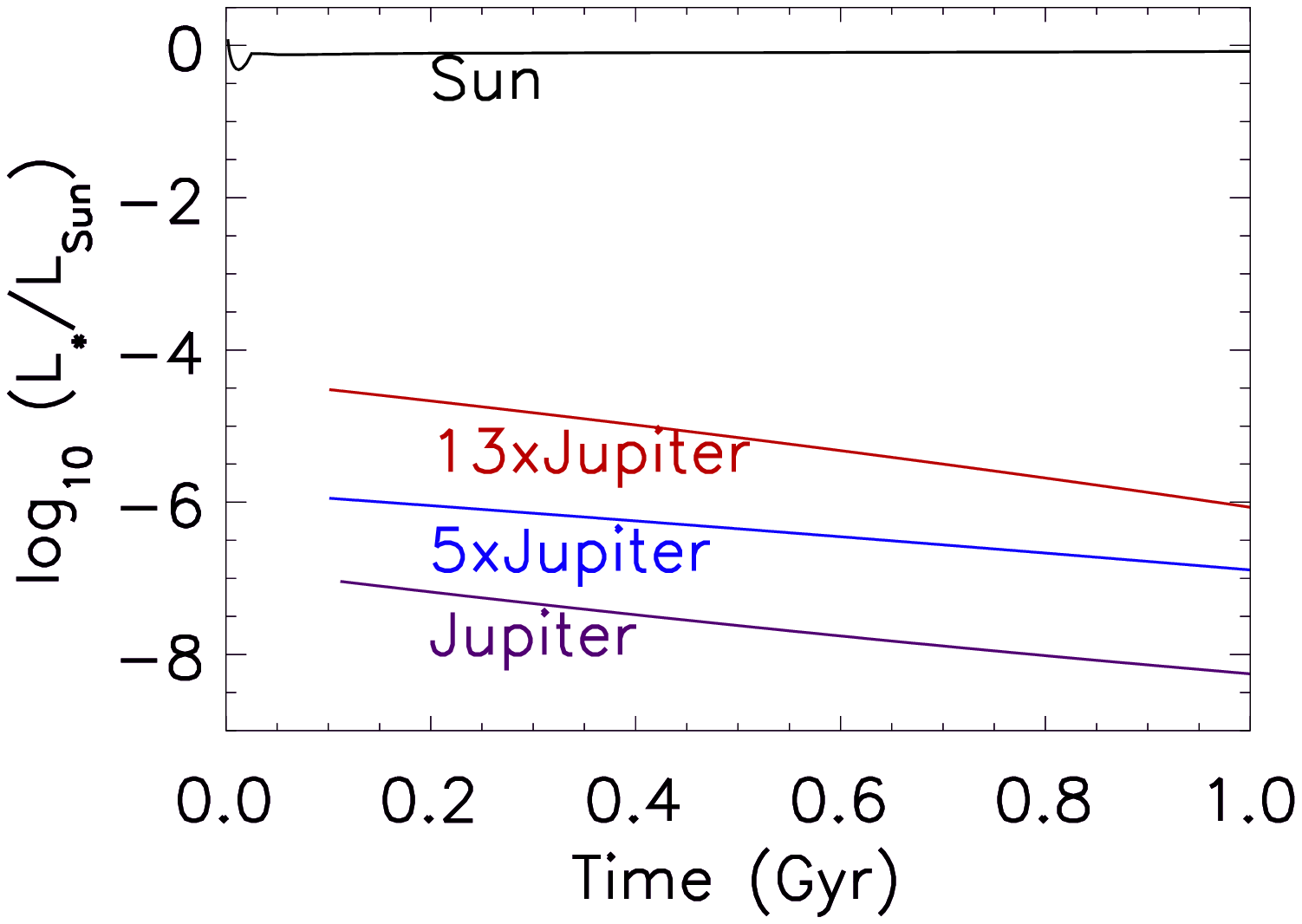}}
  \vspace{-0.4cm}
  \caption{Luminosity evolution of a Sun-like star \citep[according to][]{1998A&A...337..403B} and of three giant planets with $13$, $5$, and $1\,M_\mathrm{J}$ \citep[following][]{2003A&A...402..701B}.}
  \label{fig:cooling}
\end{figure}

To parametrize the luminosities of the star and the cooling planet, we use the cooling tracks from \citet{1998A&A...337..403B} and \citet{2003A&A...402..701B}, respectively. Figure~\ref{fig:cooling} shows the evolution of the luminosity for a Sun-like star and the luminosity of three giant planets with $13$, $5$, and $1\,M_\mathrm{J}$. Note the logarithmic scale: at an age of $0.2$ billion years (Gyr), a Jupiter-mass planet emits about 0.3\,\% the amount of radiation of a $13\,M_\mathrm{J}$ planet. These giant planets models assume arbitrarily large initial temperatures and radii. Yet, to actually assess the luminosity evolution of a giant exoplanet during the first $\approx500$\,Myr, its age, mass, and luminosity would need to be known \citep{2013A&A...558A.113M}. Our study is by necessity illustrative. Once actual exomoons are discovered more realistic giant planet models may be more appropriate.

\subsection{Tidal heating in exomoons}

Tidal heating has been identified as a possible key source for a moon's energy flux budget (see Sect.~\ref{sec:introduction}). Hence, we will also explore the combined effects of stellar and planetary illumination on extrasolar satellites plus the contribution of tidal heating.

While tidal theories were initially developed to describe the tidal flexing of rigid bodies in the solar system \citep[such as the Moon and Jupiter's satellite Io, see][]{1879_Darwin,1880RSPT..171..713D,1979Sci...203..892P,1988Icar...75..187S}, the detection of bloated giant planets in close circumstellar orbits has triggered new efforts on realistic tidal models for gaseous objects. Nowadays, various approaches exist to parametrize tidal heating, and two main realms for bodies with an equilibrium tide \citep{1977A&A....57..383Z} have emerged: constant-phase-lag (CPL) models \citep[typically applied to rigid bodies, see][]{2008CeMDA.101..171F,2009ApJ...698L..42G}, and the constant-time-lag (CTL) models \citep[typically applied to gaseous bodies, see][]{1981A&A....99..126H,2010A&A...516A..64L}.

We here make use of the CPL model of \citet{2008CeMDA.101..171F} to estimate the tidal surface heating $F_\mathrm{t}$ on our hypothetical exomoons. This model includes tidal heating from both circularization (up to second order in eccentricity) and tilt erosion \citep[that is, tidally-induced changes in obliquity $\psi_\mathrm{s}$,][]{2011A&A...528A..27H}. For simplicity, we assume that $F_\mathrm{t}$ distributes evenly over the satellite's surface, although observations of Io, Titan, and Enceladus suggest that tidal heat can be episodic and heat would leave a planetary body through hot spots \citep{1986Icar...66..341O,2000Sci...288.1198S,2005Natur.435..786S,Spencer10032006,Porco10032006,2008Icar..196..642T,2012Icar..219..655B}, that is, volcanoes or cryovolcanoes. The obliquity of a satellite in an orbit similar to Jupiter's Galilean moons and Saturn's moon Titan, with orbital periods $\lesssim16$\,d, is eroded in much less than 1\,Gyr. As in \citet{2013AsBio..13...18H}, we thus assume $\psi_\mathrm{s}=0$. We treat the moon's orbit to have an instantaneous eccentricity $e_\mathrm{ps}$. Tidal heating from circularization implies that $e_\mathrm{ps}$ approaches zero, but it can be forced by stellar, planetary, or even satellite perturbations to remain non-zero. Alternatively, it can be the remainder of an extremely large initial eccentricity, or be caused by a recent impact. Note also that Titan's orbital eccentricity of roughly 0.0288 is still not understood by any of these processes \citep{1995Icar..115..278S}.

\subsection{The runaway greenhouse effect}

To assess whether our hypothetical satellites would be habitable, we compare the sum of total illumination $F_\mathrm{i}$ (Eq.~\ref{eq:F_i}) and the globally averaged tidal heat $F_\mathrm{t}$ to the critical flux for a runaway greenhouse, $F_\mathrm{crit}$. The latter value is given by the semi-analytical approach of \citet[][Eq. 4.94 therein]{2010ppc..book.....P} \citep[see also Eq.~1 in][]{2013AsBio..13...18H}, which predicts the initiation of the runaway greenhouse effect based on the maximum possible outgoing longwave radiation from a planetary body with an atmosphere saturated in water vapor.

We study two hypothetical satellites. In the first case of a rocky Earth-mass, Earth-sized moon, we obtain a critical flux of $F_\mathrm{crit}=295\,\mathrm{W/m}^2$ for the onset of the runaway greenhouse effect. In the second case, we consider an icy moon of $0.25\,M_\oplus$\footnote{This mass corresponds to about ten times the mass of Ganymede and constitutes roughly the detection limit of \textit{Kepler} \citep{2009MNRAS.400..398K}.} and an ice-to-mass fraction of 25\,\%. Using the structure models of \citet{2007ApJ...659.1661F}, we deduce the satellite radius (0.805 times the radius of Earth) and surface gravity ($3.75\,\mathrm{m/s}^2$) and finally obtain $F_\mathrm{crit}=266\,\mathrm{W/m}^2$ for this Super-Ganymede. If one of our test moons undergoes a total flux that is beyond its critical limit, then it can be considered temporarily uninhabitable.

\subsection{An exomoon menagerie}
\label{sub:menagerie}

With decreasing distance from the planet, irradiation from the planet and tidal forces on the moon will increase. While an Earth-sized satellite in a wide circumplanetary orbit may essentially behave like a freely rotating planet with only the star as a relevant light source, moons in close orbits will receive substantial irradiation from the planet -- at the same time be subject to eclipses -- and eventually undergo enormous tidal heating.

To illustrate combined effects of illumination from the planet and tidal heat as a function of distance to the planet, we introduce an exomoon menagerie \citep{2013AsBio..13..279B}. It consists of the following specimen (colors in brackets refer to Figs.~\ref{fig:menagerie_example} and \ref{fig:menagerie}):

\begin{itemize}
\item Tidal Venus (red):\\
$F_\mathrm{t} \ {\geq} \ F_\mathrm{crit}$ \citep{2013AsBio..13..225B}\\
\item Tidal-Illumination Venus (orange):\\
$F_\mathrm{i}<F_\mathrm{crit} \ \wedge \ F_\mathrm{t}<F_\mathrm{crit} \ \wedge \ F_\mathrm{i}+F_\mathrm{t} \ {\geq} \ F_\mathrm{crit}$\\
\item Super-Io \citep[hypothesized by][yellow]{2008ApJ...681.1631J}:\\
$F_\mathrm{t}>2\,\mathrm{W/m}^2 \ \wedge \ F_\mathrm{i}+F_\mathrm{t}<F_\mathrm{crit}$\\
\item Tidal Earth (blue):\\
$0.04\,\mathrm{W/m}^2<F_\mathrm{t}<2\,\mathrm{W/m}^2 \ \wedge \ F_\mathrm{i}+F_\mathrm{t}<F_\mathrm{crit}$\\
\item Earth-like (green):\\
$F_\mathrm{t}<0.04\,\mathrm{W/m}^2$ and within the stellar HZ
\end{itemize}

Among these states, a Tidal Venus and a Tidal-Insolation Venus are uninhabitable, while a Super-Io, a Tidal Earth, and an Earth-like moon could be habitable. The 2 and $0.04\,\mathrm{W/m}^2$ limits are taken from examples in the solar system, where Io's extensive volcanism coincides with an endogenic surface flux of roughly $2\,\mathrm{W/m}^2$ \citep{2000Sci...288.1198S}. \citet{1997Natur.385..234W} estimated that tectonic activity on Mars came to an end when its outgoing energy flux through the surface fell below $0.04\,\mathrm{W/m}^2$.

For our menageries, we consider the rocky Earth-type moon orbiting a giant planet with a mass 13 times that of Jupiter. We investigate planet-moon binaries at two different distances to a Sun-like star. In one configuration, we will assume that the planet-moon duet orbits a Sun-like host at 1\,AU. In this scenario, the planet-moon system is close to the inner edge of the stellar HZ \citep{2013ApJ...765..131K}. In a second setup, the binary is assumed to orbit the star at a distance equivalent to $0.331\,S_\mathrm{eff,\odot}$, which is the average of the maximum greenhouse and the early Mars limits computed by \citet{2013ApJ...765..131K}. In this second configuration, hence, the binary is assumed at a distance of $1.738$\,AU from a Sun-like host star, that is, at the outer edge of the stellar HZ.

To explore the effect of tidal heating, we consider four different orbital eccentricities of the planet-moon orbit: $e_\mathrm{ps}\in\{10^{-4},10^{-3},10^{-2},10^{-1}\}$. For the Tidal Venus and the Tidal-Illumination Venus satellites, we assume a tidal quality factor $Q_\mathrm{s}=100$, while we use $Q_\mathrm{s}=10$ for the others. Our choice of larger $Q_\mathrm{s}$, corresponding to lower dissipation rates, for the Tidal Venus and the Tidal-Illumination Venus moons is motivated by the effect we are interested in, namely volcanism, which we assume independent of tidal dissipation in a possible ocean. Estimates for the tidal dissipation in dry solar system objects yields $Q_\mathrm{s}\approx100$ \citep{1966Icar....5..375G}. On the other moons, tidal heating is relatively weak, and we are mostly concerned with the runaway greenhouse effect, which depends on the surface energy flux. Hence, dissipation in the ocean is crucial, and because Earth's dissipation constant is near $10$, we choose the same value for moons with moderate and weak tidal heating. Ultimately, we consider all these constellations at three different epochs, namely, at ages of 100, 500, and 1000\,Myr.

In total, two stellar distances of the planet-moon duet, four orbital eccentricities of the planet-moon system, and three epochs yield 24 combinations, that is, our circumplanetary exomoon menageries.

\section{Results}
\label{sec:results}

\subsection{Evolution of stellar and planetary illumination}

\begin{figure}[t]
\vspace{-0.0cm}
  \centering
  \scalebox{0.49}{\includegraphics{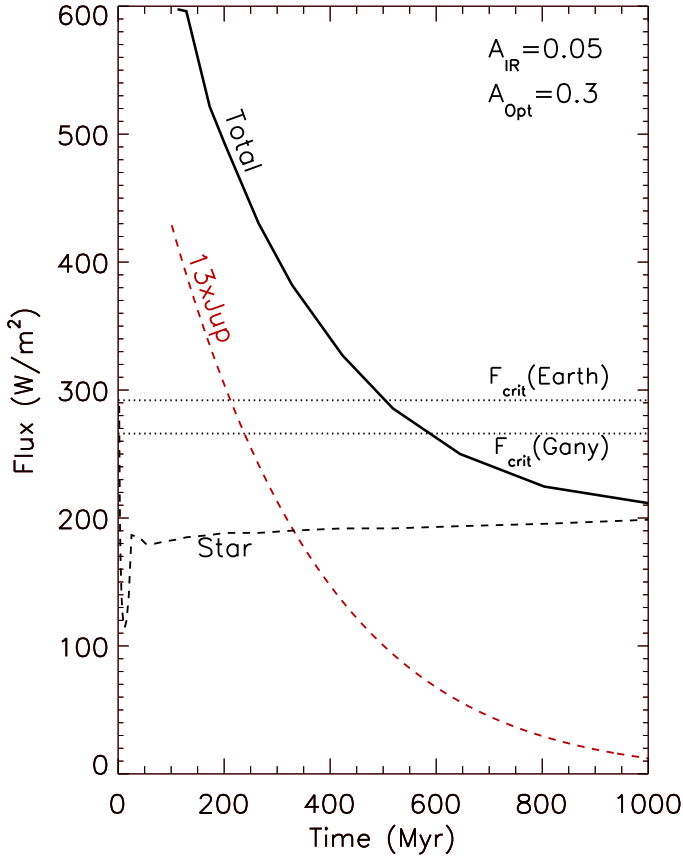}}
  \vspace{-0.1cm}
  \caption{The total illumination $F_\mathrm{i}$ absorbed by a moon (thick black line) is composed of the absorbed flux from the star (black dashed line) and from its host planet (red dashed line). The critical values for an Earth-type and a Super-Ganymede moon to enter the runaway greenhouse effect ($F_\mathrm{crit}$) are indicated by dotted lines at $295$ and $266\,\mathrm{W/m}^2$, respectively. Both moons orbit a $13\,M_\mathrm{J}$ planet at a distance of $10\,R_\mathrm{J}$, at $1$\,AU from a Sun-like host star. Illumination from the planet alone can trigger a runaway greenhouse effect for the first $\approx200$\,Myr.}
  \label{fig:flux}
\end{figure}

In Fig.~\ref{fig:flux}, we show how absorbed stellar flux (dashed black line), illumination from the planet (dashed red line), and total illumination (solid black line) evolve for a moon at $a_\mathrm{ps}=10\,R_\mathrm{J}$ from a $13\,M_\mathrm{J}$ planet.\footnote{For comparison, Io, Europa, Ganymede, and Callisto orbit Jupiter at approximately 6.1, 9.7, 15.5, and 27.2 Jupiter radii. In-situ formation of moons occurs mostly between roughly 5 and $30\,R_\mathrm{p}$, from planets the mass of Saturn up to planets with the 10-fold mass of Jupiter \citep{2010ApJ...714.1052S,HellerReview}.} At that distance, irradiation from the planet alone can drive a runaway greenhouse for about $200$\,Myr on both the Earth-type moon and the Super-Ganymede. What is more, when we include stellar irradiation from a Sun-like star at a distance of 1\,AU, then the total illumination at $10\,R_\mathrm{J}$ from the planet is above the runaway greenhouse limit of an Earth-like satellite for about $500$\,Myr. Our hypothetical Super-Ganymede would be in a runaway greenhouse state for roughly $600$\,Myr.

At a distance of $15\,R_\mathrm{J}$, flux from the planet would be $(10/15)^2=0.\bar{4}$ times the red dashed line shown in Fig.~\ref{fig:flux}. After 200\,Myr, illumination from the planet would still be $0.\bar{4}\times310\,\mathrm{W/m}^2$, and with the additional $190\,\mathrm{W/m}^2$ absorbed from the star, the total irradiation would still sum up to about $328\,\mathrm{W/m}^2$ at an age of 200\,Myr. Consequently, even at a distance similar to that of Ganymede from Jupiter, an Earth-sized moon could undergo a runaway greenhouse effect around a $13\,M_\mathrm{J}$ planet over several hundred million years. At a distance of $20\,R_\mathrm{J}$, total illumination would be $268\,\mathrm{W/m}^2$ after 200\,Myr, and our Super-Ganymede test moon would still be uninhabitable. Clearly, thermal irradiation from a super-Jupiter host planet can have a major effect on the habitability of its moons.

\subsection{Evolution of illumination and tidal heating}

\begin{figure*}[t]
\centering
\scalebox{0.51}{\includegraphics{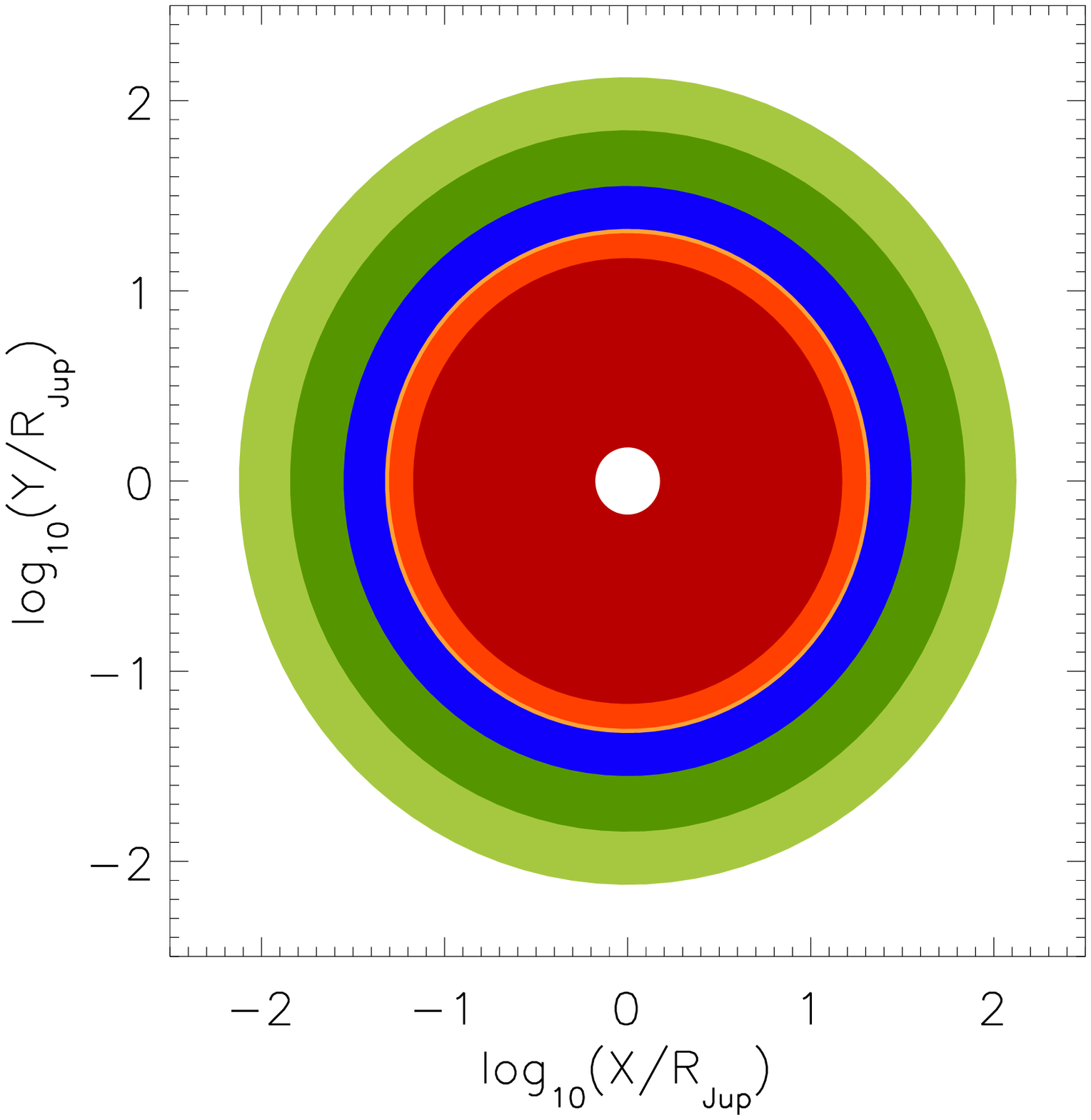}}
\hspace{1.0cm}
\scalebox{0.51}{\includegraphics{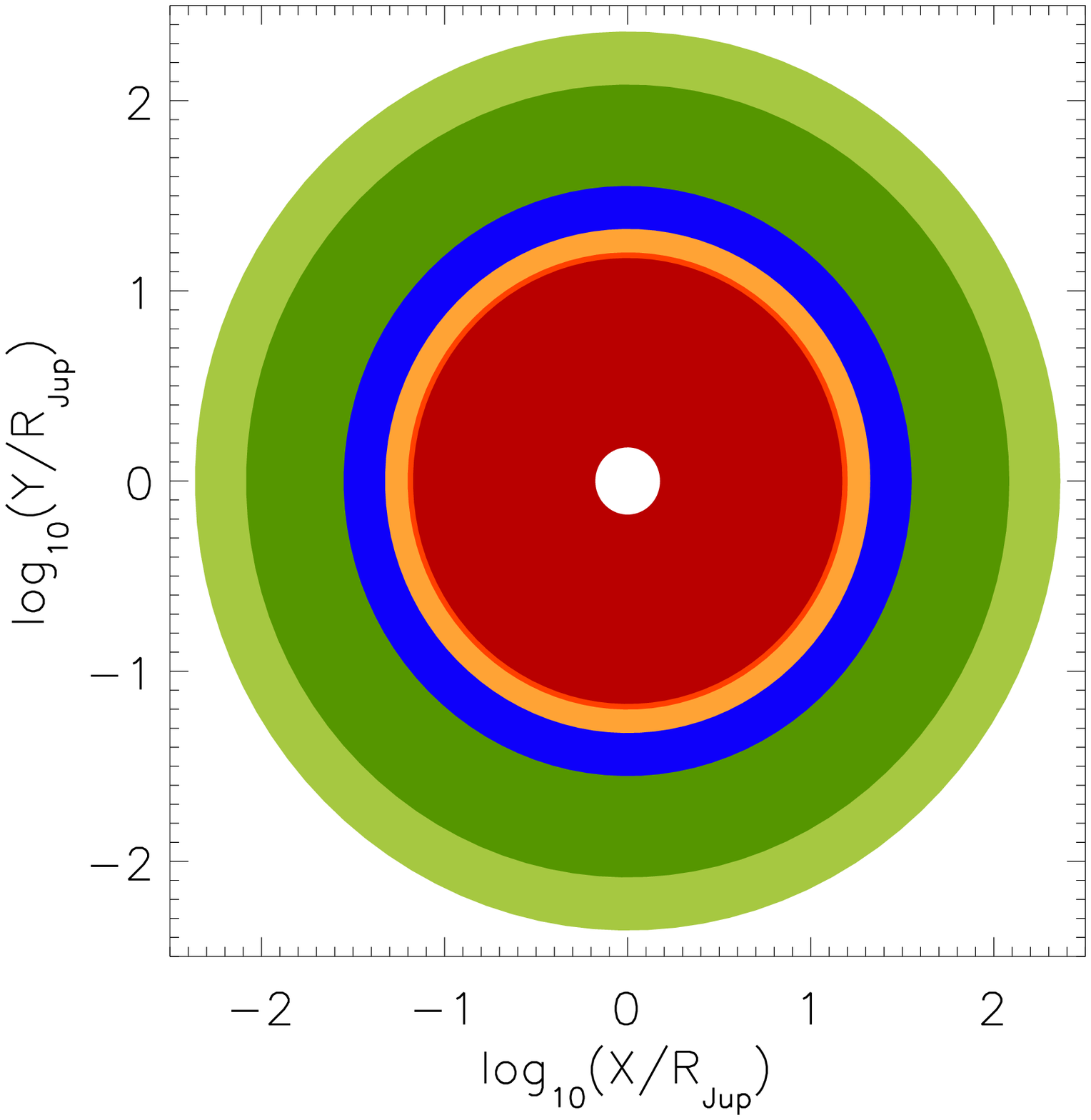}} \\
\vspace{0.5cm}
\scalebox{0.51}{\includegraphics{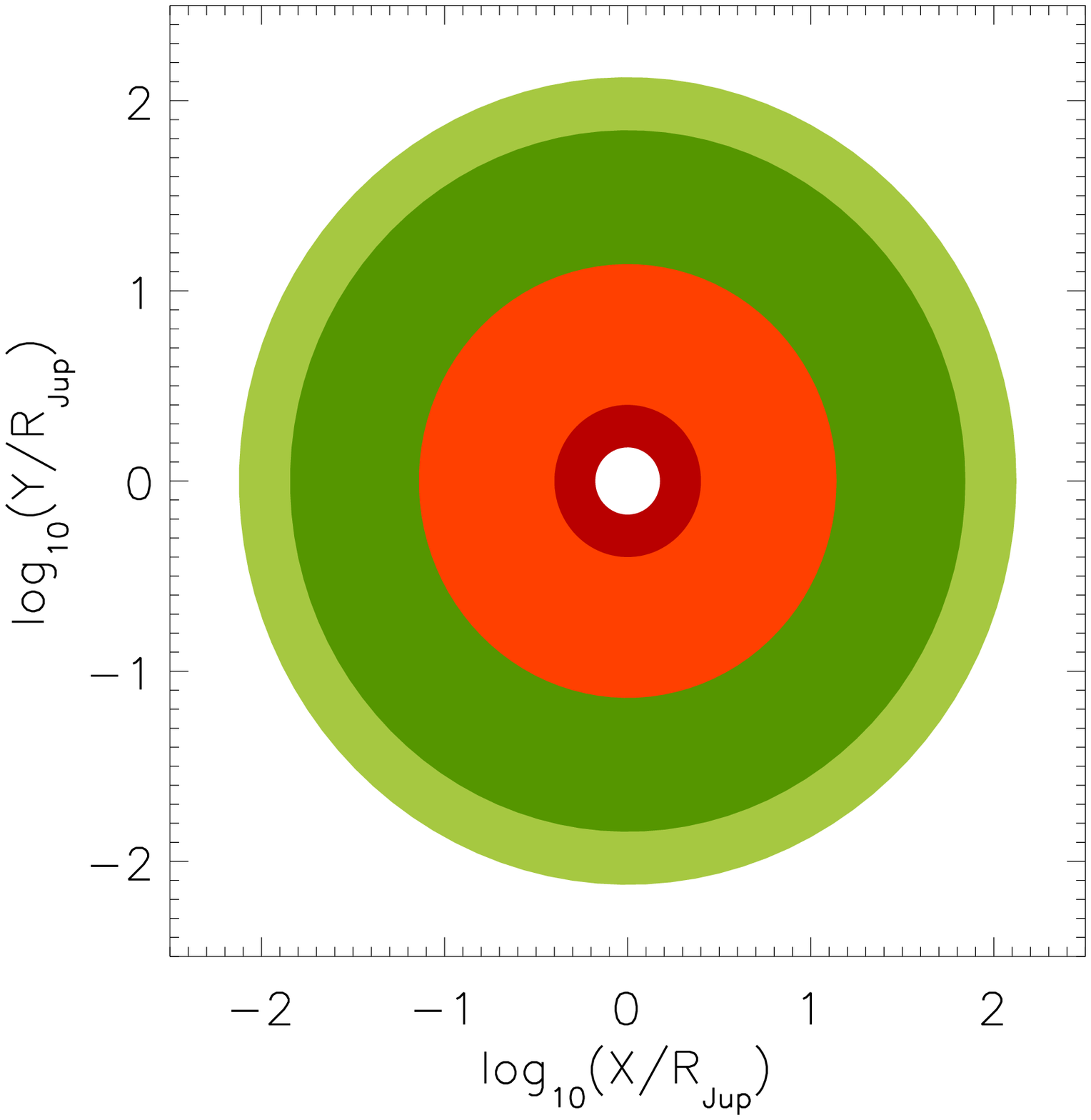}}
\hspace{1.0cm}
\scalebox{0.51}{\includegraphics{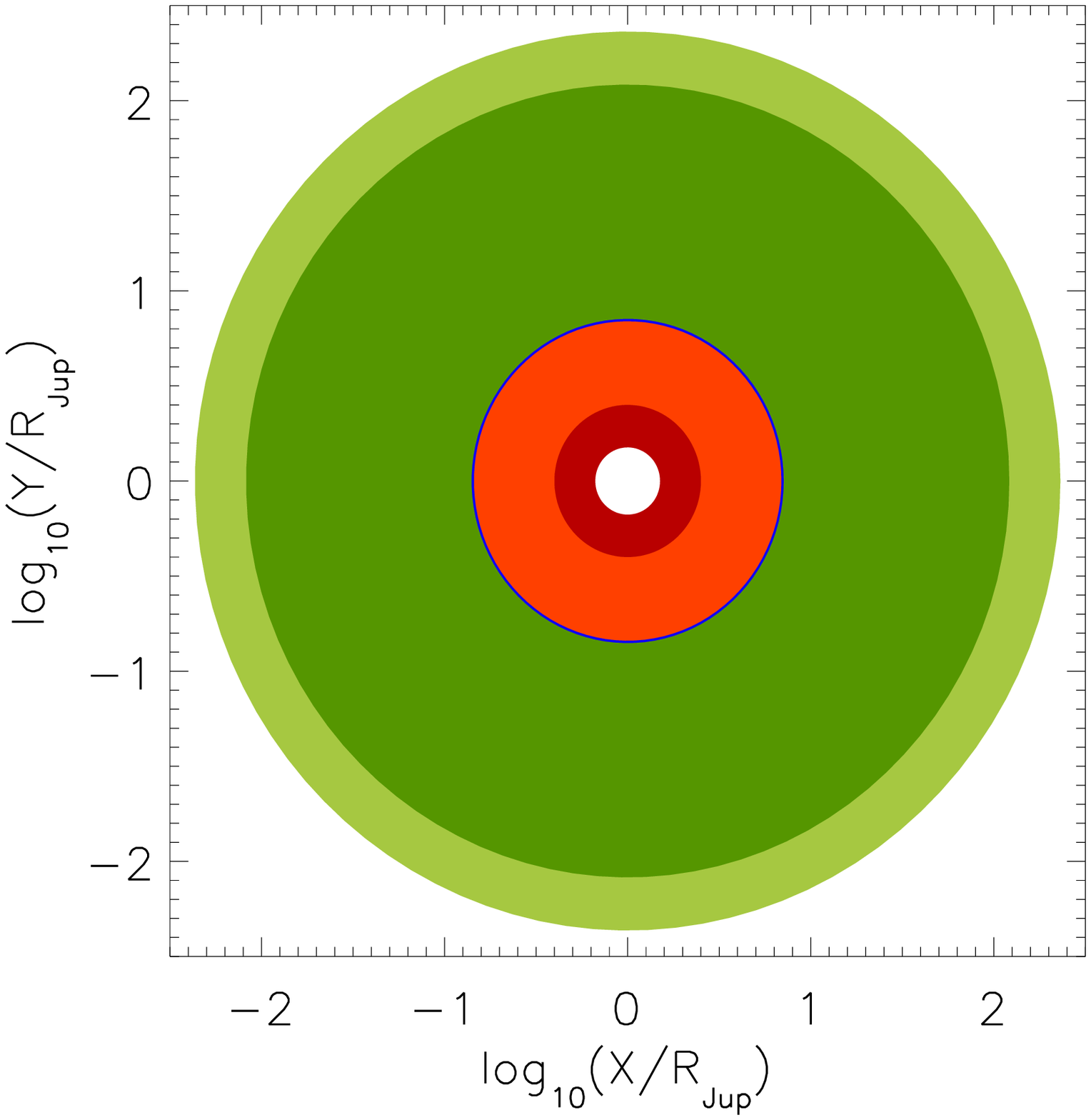}}
\caption{Circumplanetary exomoon menageries for Earth-sized satellites around a 13 Jupiter-mass host planet at an age of $500$\,Myr. In each panel, the planet's position is at (0,0), and distances are shown on a logarithmic scale. In the left panels, the planet-moon binary orbits at a distance of $1$\,AU from a Sun-like star; in the right panels, the binary is at the outer edge of the stellar HZ at $1.738$\,AU. In the upper two panels, $e_\mathrm{ps}=10^{-1}$; in the lower two panels, $e_\mathrm{ps}=10^{-4}$. Starting from the planet in the center, the white circle visualizes the Roche radius, and the exomoon types correspond to Tidal Venus (red), Tidal-Illumination Venus (orange), Super-Io (yellow), Tidal Earth (blue), and Earth-like (green) states (see Sect.~\ref{sub:menagerie} for details). Dark green depicts the Hill sphere of prograde Earth-like moons, light green for retrograde Earth-like moons. Note the larger Hill radii at the outer edge of the HZ (right panels)!}
\label{fig:menagerie_example}
\end{figure*}

\begin{figure*}[t]
\vspace{.97cm}
  \scalebox{0.434}{\includegraphics[angle=90]{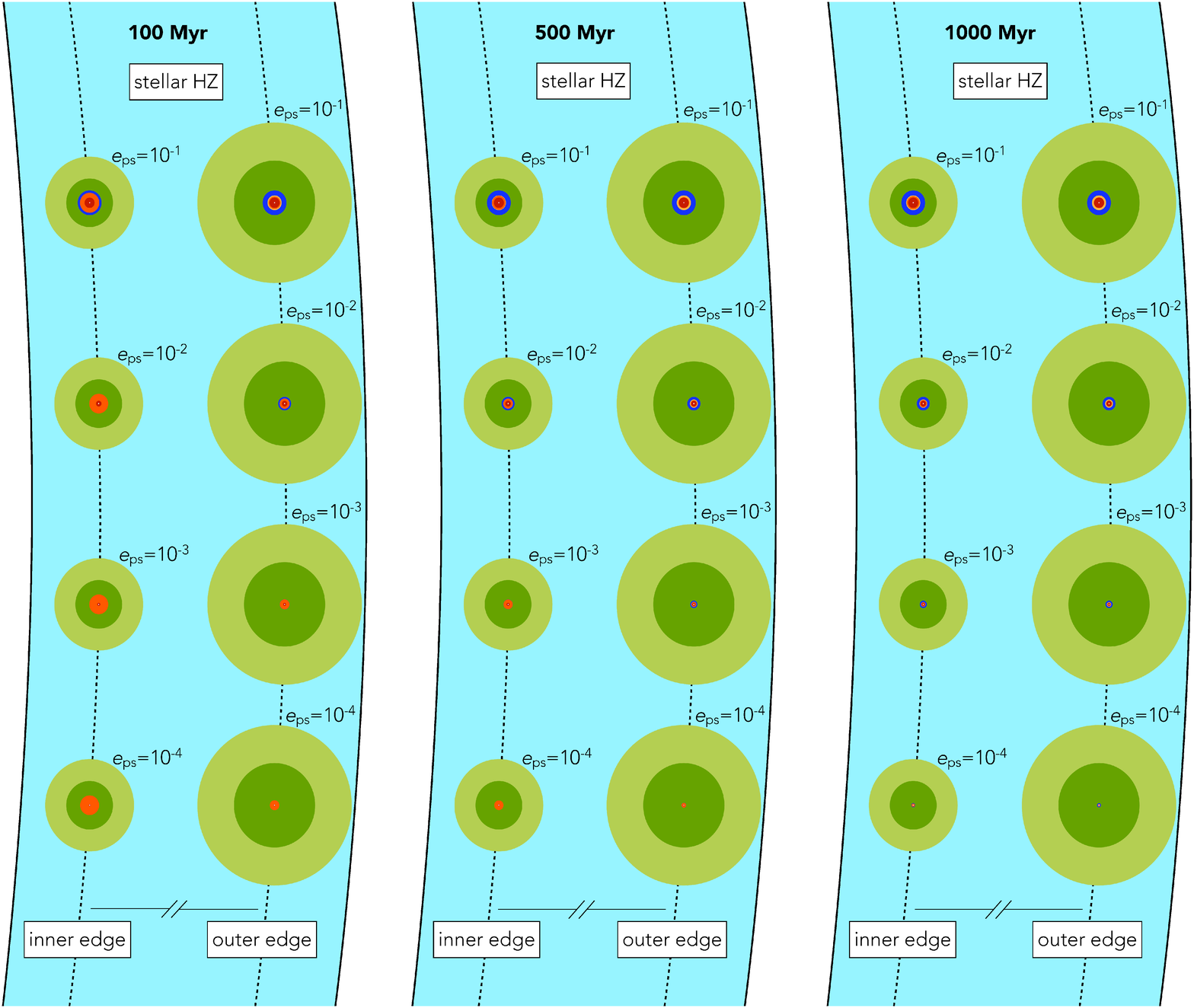}}
  \caption{Exomoon menageries at 100 (left), 500 (center), and 1000\,Myr (right) for a hypothetical Earth-like satellite around a $13\,M_\mathrm{J}$ planet. For each epoch, one suite of menageries is at the inner edge of the HZ (left) and another one at the outer edge of the HZ around a Sun-like star. Four different orbital eccentricities $e_\mathrm{ps}$ of the satellite are indicated. Distances from the central planet in each menagerie are on a linear scale, and absolute values can be estimated by comparison with the four examples shown on a logarithmic scale in Fig.~\ref{fig:menagerie_example}. The Hill sphere for prograde moons (boundary between dark and light green) is at $70\,R_\mathrm{J}$ at the inner HZ boundary and at $121\,R_\mathrm{J}$ at the outer HZ edge.}
  \label{fig:menagerie}
\end{figure*}

In Fig.~\ref{fig:menagerie_example}, we show four examples for our circumplanetary exomoon menageries. Abscissae and ordinates denote the distance to the planet, which is chosen to be located at the center at (0,0) [note the logarithmic scale!]. Colors illustrate a Tidal Venus (red), Tidal-Illumination Venus (orange), Super-Io (yellow), Tidal Earth (blue), and an Earth-like (green) state of an Earth-sized exomoon (for details see Sect.~\ref{sub:menagerie}). Light green depicts the Hill radius for retrograde moons, dark green for a prograde moons. In all panels, the planet-moon binary has an age of 500\,Myr. The two panels in the left column show planet-moon systems at the inner edge of the stellar HZ; the two panels to the right show the same systems at the outer edge of the HZ. In the top panels, $e_\mathrm{ps}=10^{-1}$ and tidal heating is strong; in the bottom panels $e_\mathrm{ps}=10^{-4}$ and tidal heating is weak.

The outermost stable satellite orbit is only a fraction of the planet's Hill radius and depends, amongst others, on the distance of the planet to the star \citep{2006MNRAS.373.1227D}: the closer the star, the smaller the planet's Hill radius. This is why the green circles are smaller at the inner edge of the HZ (left panels). In this particular constellation, the boundary between the dark and the light green circles, that is, the Hill radii for moons in prograde orbits, is at $70\,R_\mathrm{J}$, while the outermost stable orbit for retrograde moons is at $132\,R_\mathrm{J}$. At the outer edge of the HZ (right panels), these boundaries are 121 at and $231\,R_\mathrm{J}$, respectively.

A comparison between the top and bottom panels shows that tidal heating, triggered by the substantial eccentricity in the upper plots, can have a dramatic effect on the circumplanetary, astrophysical conditions. The Tidal Earth state (blue) in a highly eccentric orbit (upper panels) can be maintained between roughly 21 and $36\,R_\mathrm{J}$ both at the inner and the outer edge of the stellar HZ.\footnote{Tidal heating in the moon does not depend on the stellar distance.} But this state does not exist in the extremely low eccentricity configuration at the inner HZ edge (lower left panel). In the latter constellation, the region of moderate tidal heating is inside the circumplanetary sphere in which stellar illumination plus illumination from the planet are strong enough to trigger a runaway greenhouse effect on the moon (red and orange circles for a Tidal Venus and Tidal-Illumination Venus state, respectively). In the lower right panel, a very thin rim of a Tidal Earth state exists at roughly $8\,R_\mathrm{J}$.

Each of these four exomoon menageries has its own circumplanetary death zone, that is, a range of orbits in which an Earth-like moon would be in a Tidal Venus or Tidal-Illumination Venus state. In \citet{2013AsBio..13...18H}, we termed the outermost orbit, which would just result in an uninhabitable satellite, the ``habitable edge''. Inspection of Fig.~\ref{fig:menagerie_example} yields that the habitable edges for $e_\mathrm{ps}=10^{-1}$ (top) and $e_\mathrm{ps}=10^{-4}$ (bottom) are located at around 20 and $12\,R_\mathrm{J}$ with the planet-moon binary at the inner edge of the stellar HZ (left column), and at 15 and $8\,R_\mathrm{J}$ with the planet-moon pair orbiting at the outer edge of the HZ (right column, top and bottom panels), respectively.

Note that for moons with surface gravities lower than that of Earth, the critical energy flux for a runaway greenhouse effect would be smaller. Although tidal surface heating in the satellite would also be smaller, as it is proportional to the satellite's radius cubed, the habitable edge would be even farther away from the planet \citep[see Fig.~10 in][]{2013AsBio..13...18H}. For Mars- to Earth-sized moons, and in particular for our Super-Ganymede hypothetical satellite, the habitable edges would be even farther away from the planet than depicted in Fig.~\ref{fig:menagerie_example}.

Figure~\ref{fig:menagerie}, finally, shows our whole model grid of 24 exomoon menageries. The four examples from Fig.~\ref{fig:menagerie_example} can be found at the very top and the very bottom of the diagram in the center -- though now on a linear rather than on a logarithmic scale. The left, center, and right graphics show a range of exomoon menageries at ages of 100, 500, and 1000\,Myr, respectively. In all three epochs, highly eccentric exomoon orbits are shown at the top ($e_\mathrm{ps}=10^{-1}$), almost circular orbits at the bottom ($e_\mathrm{ps}=10^{-4}$). The four circumplanetary circles at the left illustrate menageries with the planet-moon system assumed at the inner edge of the HZ, while the four menageries at the right visualize the planet-moon duet at the outer edge of the HZ.

In the highly eccentric cases at the top, tidal heating dominates the circumplanetary orbital conditions for exomoon habitability. In the system's youth at 100\,Myr, the circumplanetary habitable edge is as far as $29\,R_\mathrm{J}$ (upper left panel, at the inner edge of the HZ) or $18\,R_\mathrm{J}$ (right panel in the leftmost diagram, outer edge of the HZ) around the planet. The region for Tidal Earth moons spans from roughly 29 to $36\,R_\mathrm{J}$ at the inner edge of the HZ and from 21 to $36\,R_\mathrm{J}$ at the outer edge of the stellar HZ. At an age of 1000\,Myr (diagram at the right), when illumination from the planet has decreased by more than one order of magnitude (see red line in Fig.~\ref{fig:cooling}), the habitable edge has moved inward to roughly $19\,R_\mathrm{J}$ at the inner edge and $16\,R_\mathrm{J}$ at the outer edge of the HZ, while the Tidal Earth state is between 21 to $36\,R_\mathrm{J}$ in both the inner and outer HZ edge cases.

In the low-eccentricity scenarios at the bottom, tidal heating has a negligible effect, and thermal flux from the planet dominates the evolution of the circumplanetary conditions. At 100\,Myr, the Tidal-Illumination Venus state, which is just inside the planetary habitable edge, ranges out to 28$\,R_\mathrm{J}$ at the inner edge of the HZ (lower left) and to 14$\,R_\mathrm{J}$ at the outer edge of the HZ (lower right menagerie in the left-most diagram). At an age of 1000\,Myr, those values have decreased to 5 and 2.5$\,R_\mathrm{J}$ for the planet-moon duet at the inner and outer HZ boundaries, respectively (bottom panels in the rightmost sketch). Moreover, at 1000\,Myr illumination from the planet has become weak enough that moons with substantial tidal heating in these low-eccentricity scenarios could exist between 5 and 6$\,R_\mathrm{J}$ (Tidal Earth) at the inner HZ edge or between 2.5 and 6$\,R_\mathrm{J}$ (Super-Io and Tidal Earth) at the outer HZ boundary.

\section{Conclusions}
\label{sec:conclusions}

Young and hot giant planets can illuminate their potentially habitable, Earth-sized moons strong enough to make them uninhabitable for several hundred million years. Based on the planetary evolution models of \citet{2003A&A...402..701B}, thermal irradiation from a $13\,M_\mathrm{J}$ planet on an Earth-sized moon at a distance of $10\,R_\mathrm{J}$ can trigger a runaway greenhouse effect for about $200$\,Myr. The total flux of Sun-like irradiation at 1\,AU plus thermal flux from a $13\,M_\mathrm{J}$ planet will force Earth-sized moons at 10 or $15\,R_\mathrm{J}$ into a runaway greenhouse for 500 or more than 200\,Myr, respectively. A Super-Ganymede moon $0.25$ times the mass of Earth and with an ice-to-mass fraction of $25\,\%$ would undergo a runaway greenhouse effect for longer periods at the same orbital distances or, equivalently, for the same periods at larger separations from the planet. Even at a distance of $20\,R_\mathrm{J}$ from a $13\,M_\mathrm{J}$ host giant, it would be subject to a runaway greenhouse effect for about 200\,Myr, if it receives an early-Earth-like illumination from a young Sun-like star. In all these cases, the moons could lose substantial amounts of hydrogen and, consequently, of water. Such exomoons would be temporarily uninhabitable and, perhaps, uninhabitable forever.

If tidal heating is included, the danger for an exomoon to undergo a runaway greenhouse effect increases. The habitable edge around young giant planets, at an age of roughly 100\,Myr and with a mass 13 times that of Jupiter, can extend out to about $30\,R_\mathrm{J}$ for moons in highly eccentric orbits ($e_\mathrm{ps}\approx0.1$). That distance encompasses the orbits of Io, Europa, Ganymede, and Callisto around Jupiter. Beyond the effects we considered here, that is, stellar irradiation, thermal irradiation from the planet, and tidal heating, other heat sources in moons may exist. Consideration of primordial thermal energy (or ``sensible heat''), radioactive decay, and latent heat from solidification inside an exomoon would increase the radii of the circumplanetary menageries and push the habitable edge even further away from the planet.

Our estimates for the instantaneous habitability of exomoons should be regarded as conservative because (\textit{i.}) the minimum separation for an Earth-like moon from its giant host planet to be habitable could be even larger than we predict. This is because a giant planet's luminosity at 1\,AU from a Sun-like star can decrease more slowly than in the \citet{2003A&A...402..701B} models used in our study \citep{2007ApJ...659.1661F}. (\textit{ii.}) The runaway greenhouse limit, which we used here to assess instantaneous habitability, is a conservative approach itself. A moon with a total energy flux well below the runaway greenhouse limit may be uninhabitable as its surface is simply too hot. It could be caught in a moist greenhouse state with surface temperatures up to the critical point of water, that is, 647\,K if an Earth-like inventory of water and surface H$_2$O pressure are assumed \citep{1988Icar...74..472K}. (\textit{iii.}) Moons with surface gravities smaller than that of Earth will have a critical energy flux for the runaway greenhouse effect that is also smaller than Earth's critical flux. Hence, their corresponding Tidal Venus and Tidal-Illumination Venus states would reach out to wider orbits than those depicted in Figs.~\ref{fig:menagerie_example} and \ref{fig:menagerie}. (\textit{iv.}) Tidal heating could be stronger than the values we derived with the \citet{2008CeMDA.101..171F} CPL tidal model. As shown in \citet{2011A&A...528A..27H}, the CTL theory of \citet{2010A&A...516A..64L} includes terms of higher orders in eccentricity and yields stronger tidal heating than the CPL of \citet{2008CeMDA.101..171F}. Yet, such mathematical extensions may not be physically valid \citep{2009ApJ...698L..42G}, and parametrization of a planet's or satellite's tidal response with a constant tidal quality factor $Q$ or a fixed tidal time lag $\tau$ remains uncertain \citep{2010A&A...516A..64L,2011A&A...528A..27H,2013ApJ...764...26E}.

Massive moons can bypass an early runaway greenhouse state if they form after the planet has cooled sufficiently. Possible scenarios for such a delayed formation include gravitational capture of one component of a binary planet system, the capture of Trojans, gas drag or pull-down mechanisms, moon mergers, and impacts on terrestrial planets \citep[for a review, see Sect. 2.1 in][]{2013AsBio..13...18H}. Alternatively, a desiccated moon in the stellar HZ could be re-supplied by cometary bombardment later. Reconstruction of any such event would naturally be difficult.

Once extrasolar moons will be discovered, assessments of their habitability will depend not only on their current orbital configuration and irradiation, but also on the history of stellar and planetary luminosities. Even if a Mars- to Earth-sized moon would be found about a Jupiter- or super-Jupiter-like planet at, say, 1\,AU from a Sun-like star, the moon could have lost substantial amounts of its initial water reservoir and be uninhabitable today. In the most extreme cases, strong thermal irradiation from the young, hot giant host planet could have desiccated the moon long ago by the runaway greenhouse effect.

\begin{acknowledgements}
The referee reports of Jim Kasting, Nader Haghighipour, and an anonymous reviewer significantly improved the quality of this study. We thank Jorge I. Zuluaga for additional comments on the manuscript and Jean Schneider for technical support. Ren{\'e} Heller is funded by the Canadian Astrobiology Training Program and a member of the Origins Institute at McMaster University. Rory Barnes acknowledges support from NSF grant AST-1108882 and the NASA Astrobiology Institute's Virtual Planetary Laboratory lead team under cooperative agreement No. NNH05ZDA001C. This work has made use of NASA's Astrophysics Data System Bibliographic Services and of Jean SchneiderÕs exoplanet database \url{www.exoplanet.eu}.
\end{acknowledgements}

\bibliographystyle{aa} 
\bibliography{2013-01_HellerBarnes--Exomoon_Inplanation}

\end{document}